\documentstyle[preprint,aps]{revtex}

\newcommand{\be}{\begin{eqnarray}}
\newcommand{\ee}{\end{eqnarray}}
\newcommand{\Del}{\Delta}

\begin{document}


\preprint{
KAIST-TH-98/01
}

\title{ 
$B-\bar{B}$ Mixings and Rare $B$ Decays in the Topflavour Model
}
\vspace{0.65in}

\author{
Kang Young ~Lee\thanks{kylee@chep5.kaist.ac.kr }
and 
Jong Chul ~Lee\thanks{jclee@chep5.kaist.ac.kr} 
}
\vspace{.5in}

\address{
Department of Physics, Korea Advanced Institute of Science and Technology\\
Taejon 305 -- 701, Korea}

\date{\today}
\maketitle

\begin{abstract}
\\
We explore the impact of topflavour model
to $B-\bar{B}$ mixings and rare decays of $B$ mesons.
By the flavour--changing neutral current interactions
of this model, the $B-\bar{B}$ mixing amplitudes
are substantially affected while it is hard to
investigate the effects on rare decays.
Violation of the unitarity of the CKM
matrix is also discussed.
We find that the bound on $|V_{td}|$ can be stronger
in this model by combining the experimental bound
of the $B-\bar{B}$ mixing and the bound from violated unitarity.

\end{abstract}


\narrowtext

\section{Introduction}

After the observation of a nonvanishing amount of
$B_d-\bar{B}_d$ mixing \cite{ua1,argus},
it has been one of the most interesting phenomena
of the $B$ meson system.
As being an flavour-changing neutral current (FCNC) process, 
the $B-\bar{B}$ mixing involves top quark in its loops 
and consequently it is sensitive to the flavour dynamics 
of the third generations
as well as being a source of the Cabbibo-Kobayashi-Maskawa (CKM)
matrix elements for the top quark and top quark mass. 
The strengths of $B_q-\bar{B}_q$ mixings are measured 
by the mass difference
\be
\Del M_q = 2 | M_{12}^{(q)}|~, 
\ee
where $M_{12}^{(q)}$ is the off--diagonal term
of the mass matrix of neutral $B_q^0$ mesons.
The present world average for measured $\Del M_d$ and the best limit 
for $\Del M_s$ are reported by \cite{deltam}
\be
\Del M_d = 0.464 \pm 0.018~~ \mbox{ps}^{-1},~~~ 
\Del M_s > 7.8~~ \mbox{ps}^{-1},
\ee
which are consistent with the Standard Model (SM) predictions 
and constrain the CKM parameters $V_{td}$ and $V_{ts}$:
\be
0.15 < \left| \frac{V_{td}}{V_{cb}} \right| < 0.34, ~~~
\left| \frac{V_{ts}}{V_{cb}} \right| > 0.6 ~.
\ee
Please see Ref. \cite{alilondon} 
to understand the experimental errors 
and theoretical uncertainties for these constraints.
This bound of $\Del M_d$ is better than that obtained 
from unitarity of the CKM matrix alone, which gives
\be
0.11 < \left| \frac{V_{td}}{V_{cb}} \right| < 0.33~.
\ee

Many models of new physics beyond the SM 
provide new contributions to $B-\bar{B}$ mixings 
which could result in altering the constraints 
on $V_{td}$ and $V_{ts}$ given in Eq. (3).
In order to be detectable, however, 
new contributions to $B-\bar{B}$ mixings
should be at least comparable to those of the SM.
This paper focuses on the model that include 
the separate SU(2) group for the third generations, 
so--called topflavour model.
This model alters the mass difference $\Del M_q$
in two distinct ways: the additional contributions
to the box diagrams and the FCNC interactions to
$Z$ and $Z'$ bosons.
Since the couplings of the left--handed third generation fermions
are different from those of the first and second generations,
the FCNC interactions appear at tree level and it renders
one of the characteristic features of the topflavour model.
We show that the new FCNC effects can be as large as that of the
box diagram of the SM.
When the $B-\bar{B}$ mixings are affected,
generically rare decay modes of $B$ mesons 
via penguin diagrams are also affected by this new physics.
Thus it is demanded to investigate 
the flavour--changing rare decays 
in the topflavour model.
The new contributions are also 
due to the FCNC interactions at tree level
as well as the additional contributions to the penguin diagrams.

This model was constrained by the LEP data 
in Ref. \cite{malkawi,nandi,lee}
and the lower bound of the mass of the additional $Z'$ boson
is predicted to be about 1.1 TeV.
We expect that the model could be also examined 
by the current data of $B-\bar{B}$ mixings 
and rare decay here.
In  this model, the CKM matrix is in general
not uniquely defined but depends upon
each bases of $U$-- type and $D$-- type quarks.
Accordingly an additional phase can be introduced in the model
to produce the new CP violating phenomena
and more parameters are needed to represent quark mixings.
In this paper, however, we choose the case 
that the weak eigenstates of $U$--type quarks are identified
as mass eigenstates for simplicity: 
$V_U = I$ and $V_D = V^0_{_{CKM}}$
respectively, where $V_{U(D)}$ represents
the unitary transformation diagonalizing
$U(D)$--type quark mass matrix.
Then we have no more phases and 
no additional CP violating phenomena.
Even in this case the CKM matrix is no more unitary.

This paper is organized as follows.
In section II, we briefly review the topflavour model.
The unitarity violation of the CKM matrix is discussed
and the FCNC is described.
The contributions to the $B-\bar{B}$ mixings are studied 
in section III and the contributions to the flavour--changing
penguin decays are examined in section IV.
We conclude in section V.

\section{The Model}
              
We study the topflavour model 
with the extended electroweak gauge group  
$SU(2)_l \times SU(2)_h \times U(1)_Y $.
The first and the second generations couple to $SU(2)_l$ 
and the third generation couples to $SU(2)_h$.
This gauge group can arise as the theory
at the intermediate scale in the pattern of gauge
symmetry breaking of noncommuting extended technicolor
(ETC) models \cite{ETC}, in which the gauge groups for
extended technicolor and for the weak interactions
do not commute.
The left--handed quarks and leptons in the first and second generations
transform as (2,1,1/3), (2,1,-1) under 
$SU(2)_l \times SU(2)_h \times U(1)_Y $,
and those in the third generation as (1,2,1/3), (1,2,-1)
while right--handed quarks and leptons transform as (1,1,2$Q$)
where $Q$ is the electric charge of fermions.
 
The covariant derivative is given by 
\begin{equation}
 D^{\mu} = \partial^{\mu} + i g_l T^{a}_{l} W^{\mu}_{la} 
          + i g_h T^{a}_{h} W^{\mu}_{ha} 
          + i g^{\prime} \frac{Y}{2} B^{\mu} , 
\end{equation}
where $ T^{a}_{l}$ and $T^{a}_{h}$ denote the $SU(2)_{(l,h)}$ 
generators and $Y$ is the $U(1)$ hypercharge generator. 
Corresponding gauge bosons are $ W^{\mu}_{la} , W^{\mu}_{ha} $
and $B^{\mu}$ with the coupling constants  
$g_{l}, g_{h}$ and $g^{\prime}$ respectively.
The gauge couplings may be written as
\begin{equation}
g_l = \frac{e}{\sin \theta \cos \phi} ,\mbox{~~~~}
g_h = \frac{e}{\sin \theta \sin \phi},\mbox{~~~~} 
g^{\prime} = \frac{e}{\cos \theta} 
\end{equation}
in terms of the weak mixing angle $\theta $ and the new mixing 
angle $\phi$ between $SU(2)_l$ and $SU(2)_h$ defined in eq. (3) below.

The symmetry breaking is accomplished  by the vacuum expectation values 
(VEV) of two scalar fields  $\Sigma$ and $\Phi$:
$ \langle \Phi \rangle = ( 0, v/\sqrt{2} )^{\dagger}$,
$ \langle \Sigma \rangle = u I$ where $I$ is $2\times2$
identity matrix.
The scalar field $\Sigma$  transforms as (2,2,0) 
under $ SU(2)_l \times SU(2)_h \times U(1) $
and we choose that $\Phi$ as (2,1,1) corresponding to the SM Higgs field.   
In the first stage, 
the scalar field $\Sigma$ gets the vacuum expectation value and
breaks $SU(2)_l \times SU(2)_h \times U(1)_Y $ down 
to $ SU(2)_{l+h} \times U(1)_Y $ at the scale $\sim u$.
The remaining symmetry is broken down to $U(1)_{em}$  
by the the VEV of $\Phi$ at the electroweak scale. 
Since the third generation fermions do not couple to Higgs fields
with this particle contents,
they should get masses via higher dimensional operators.
The different mechanism of mass generation
could be the origin of the heavy masses of the third generation.
Or it is possible to explain it
by introducing another Higgs doublet
coupled to the third generations as in  Ref. \cite{nandi}.

We demand that both $SU(2)$ interactions are perturbative
so that the value of the mixing angle $\sin \phi$
is constrained $g_{(l,h)}^2/4 \pi < 1$,
which results in $ 0.03 < \sin^2 \phi < 0.96 $.
In fact, we are interested in the region $\phi < \pi/2$
leading to $g_h> g_l$ where the third generation coupling
is stronger than those of the first and second generations.
We also assume that the first symmetry breaking scale is
much higher than the electroweak scale, 
$ v^2/u^2 \equiv \lambda \ll 1 $.

In terms of mass eigenstates of gauge bosons 
the covariant derivative can be rewritten as  
\begin{eqnarray}
D_{\mu} &=& 
\partial_{\mu} 
+ \frac{i e}{\sin \theta}  
\left[ T^{\pm}_{h} + T^{\pm}_{l} 
       + \lambda \sin^2 \phi \left( 
       \cos^2 \phi T^{\pm}_{h} - \sin^2 \phi T^{\pm}_{l} \right) 
\right] W^{\pm}_{\mu} 
\nonumber \\
&&+\frac{i e}{\sin \theta} 
\left[ \frac{\cos \phi}{\sin \phi} T^{\pm}_{h} 
       - \frac{\sin \phi}{\cos \phi} T^{\pm}_{l}
       - \lambda \sin^3 \phi \cos \phi 
         \left( T^{\pm}_{h} + T^{\pm}_{l} \right)
\right] W^{\prime \pm }_{\mu}   
\nonumber \\
&&+\frac{i e}{\sin \theta \cos \theta}  
\left[ T_{3 h} + T_{3 l} - Q \sin^2 \theta 
       + \lambda \sin^2 \phi 
       \left( \cos^2 \phi T_{3 h} - \sin^2 \phi T_{3 l} \right) 
\right] Z_{\mu}  
\nonumber \\ 
&&+\frac{i e}{\sin \theta} 
\left[ \frac{\cos \phi}{\sin \phi} T_{3 h} 
       - \frac{\sin \phi}{\cos \phi} T_{3 l} 
       - \lambda \frac{\sin^3 \phi \cos \phi}{\cos^2 \theta} 
       \left( T_{3 h} + T_{3 l} - Q \sin^2 \theta \right) 
\right] Z^{\prime}_{\mu} 
\nonumber \\
&&+ i e Q A_{\mu}  
\end{eqnarray} 
where $Q$ is the electric charge operator,
$ Q = T_{3l} + T_{3h} + Y/2 $.
The additional gauge bosons get masses such as
\be
m_{_{W'^{\pm}}}^2 = m_{_{Z'}}^2 
= m_0^2 
\left( \frac{1}{\lambda \sin^2 \phi \cos^2 \phi} + \tan^2 \phi \right)~,
\ee
while the ordinary gauge boson masses are given by
$ m_{_{W^{\pm}}}^2 
= m_0^2 ( 1- \lambda \sin^4 \phi ) 
= m_{_{Z}}^2 \cos^2 \theta$
where $m_0 = ev/(2 \sin \theta)$.

The couplings to the gauge bosons for the third generations 
are different from those of the first and second generations
and we separate the nonuniversal part
from the universal part.
First we consider the charged current:
\begin{eqnarray}
 {\cal L}^{(cc)} &=&  {\cal L}^{(cc)}_{I} + {\cal L}^{(cc)}_{3}~~,
\ee
where $ {\cal L}^{(cc)}_{I} $ denotes the universal part
and $ {\cal L}^{(cc)}_{3} $ the nonuniversal part.
We consider the unitary matrices $V_{U}$ and $V_{D}$
diagonalizing $U$--type and $D$--type quark mass matrices 
respectively.
The universal part is given by
\be
 {\cal L}_{I} &=&  {\bar U}_L {\gamma}_{\mu}
       \left[
               G_{L} W^{\mu}  + G^{\prime }_{L} W^{\prime \mu}
       \right]
         (V_U^{\dagger} V_D) D_L
       + \mbox{H.c.}
\ee
where
\begin{eqnarray}
G_L  &=&  -\frac{g}{\sqrt{2}} 
              \left( 1 - \lambda \sin^4 \phi \right) I 
\nonumber \\
G^{\prime}_L  &=& \frac{g}{\sqrt{2}} 
              \left( \tan \phi + \lambda \sin^3 \phi \cos \phi \right) I  
\nonumber \\
\end{eqnarray}
with the 3$\times$3 identity matrix $I$
and $U = (u,c,t)^{T}$ and $D = (d,s,b)^{T}$.
We define the unitary matrix 
$V_{_{CKM}}^{0} \equiv V_U^{\dagger} V_D$
which is corresponding to the
CKM matrix of the SM.
The nonuniversal part is written in terms of
mass eigenstates as:
\be
{\cal L}_3^{(cc)} &=& 
(V_{U 31}^{*} \bar{u}_L + V_{U 32}^{*} \bar{c}_L + V_{U 33}^{*} \bar{t}_L)
\nonumber \\
&& \times \gamma^{\mu} (X_L W^{+}_{\mu} +X'_L W^{\prime +}_{\mu} )
(V_{D 31} d_L + V_{D 32} s_L + V_{D 33} b_L)~,
\ee
where
\begin{eqnarray}
X_L  &=&  -\frac{g}{\sqrt{2}} 
              \lambda \sin^2 \phi \cdot 
\nonumber \\
X^{\prime}_L  &=& -\frac{g}{\sqrt{2}} 
              \left( \frac{1}{\sin \phi \cos \phi}
              \right)~.
\nonumber \\
\end{eqnarray}
Because of the existence of ${\cal L}_3$, the quark mixing matrix is
no more unitary.
Moreover it cannot be uniquely defined but depends upon the
elements of each matrices $V_U$ and $V_D$ :$\{ V_{U3j},V_{D3k}\}$.
Hence the number of parameters are doubled.
With this property, it would be very interesting to study
the model since we expect lots of new phenomena, 
e.g. additional CP violating phases.
At the same time, however, it is  too complex
and exhausting to analyze them fully.
We choose the simplest bases here: 
$V_U = I$ and $V_D = V^0_{_{CKM}}$ 
in order to avoid so tedius analysis
because we just intend to investigate 
the impact of the topflavour model.
As a matter of fact, our choice is not so arbitrary
but has a few reason in its own way.
In this basis, only three elements of $V_{td}$, $V_{ts}$, $V_{tb}$
are altered which are not experimentally examined yet.
Measured values of other six elements almost show unitarity at present.
And our basis provides the economical extension
which means that no new parameters 
do not enter the theory
for the quark mixing matrix. 
As we see below, additional contributions to
$B - \bar{B}$ mixings and rare decays contain 
the same CKM factors as appear in the SM.
Then we have the modified CKM matrix in the lagrangian:
\be
V_{_{CKM}} = V_{_{CKM}}^0 + \left( \begin{array}{ccc}
                0      &    0   &    0 \\
                0      &    0   &    0 \\
                V^0_{td} & V^0_{ts} &  V^0_{tb} \\
                            \end{array}
                          \right) \cdot \lambda~\sin^2 \phi~,
\ee
which describes the quark mixings for the charged currents
coupled to the $W^{\pm}$ bosons.
The mixing matrix for $W'^{\pm}$ bosons has the same structure
as above matrix while the model parameter $\lambda \sin^2 \phi$
is changed to $1/\sin \phi \cos \phi$.

Considering the neutral current interaction terms,
we face up to more interesting outcomes.
The nonuniversal terms bring forth the flavour--changing
neutral current interactions at tree level
which does not exist in the SM.
We discussed the FCNC effects in Ref. \cite{lee}, 
especially stressed on the lepton number violating processes.
For the quark sector, the FCNC interaction terms evolve
with the basis used above as follows
\be
{\cal L}_3^{(nc)} &=& 
(V_{D 31}^* \bar{d}_L + V_{D 32}^* \bar{s}_L + V_{D 33}^* \bar{b}_L)~,
\nonumber \\
&& \times \gamma^{\mu} (Y_L Z^{0}_{\mu} +Y'_L Z^{\prime 0}_{\mu} )
(V_{D 31} d_L + V_{D 32} s_L + V_{D 33} b_L)~,
\ee
where
\be
Y_L &=& \frac{g}{2 \cos \theta} \cdot \lambda \sin^2 \phi
\nonumber \\
Y'_L &=& \frac{g}{2} \cdot \frac{1}{ \sin \phi \cos \phi}~.
\ee
These would yield the main additional contributions to the
$B - \bar{B}$ mixings and rare decays via $Z$ and $Z'$
exchange diagrams, which will be discussed in the subsequent
sections.

\section{$ B - \bar{B}$ MIXING}

The $ B_q-\bar{B}_q$ mixings in SM are described by
the off-diagonal term $M^{(q)}_{12}$ of the mass matrix of $B_q^0$ 
\cite{smbb}:
\begin{equation}
 M^{q}_{12} = \frac{G^2_F m_{_{B_q}} \eta_{_{B_q}} m_{_{W}}^2}{6 \pi^2}
                 f^2_{B_q} B_{B_q} x_t f_2(x_t) ( V_{tq} V^*_{tb} )^2 ,
\end{equation}
where $ x_t = m_t^2/m_{_W}^2$ and
\begin{equation}
 f_2(x) = \left[ \frac{1}{4} + \frac{9}{4} \frac{1}{1-x} -
                 \frac{3}{2} \frac{1}{(1-x)^2} -
                \frac{3}{2} \frac{ x^2 \ln{x} }{(1-x)^3}
          \right]
\end{equation}
and $q =d$, $s$.
$\eta_{_{B_q}}$ denotes the QCD correction and 
$f^2_{B_q} B_{B_q}$ represents our lack of knowledge 
of the hadronic matrix elements.
Their values are quoted in Ref. \cite{alilondon}:
$\eta_{_{B_q}} = 0.55$ and $f_{B_d} \sqrt{B_{B_d}} = 200\pm40$.

There are three kinds of additional contributions 
in topflavour model to $B-\bar{B}$ mixing amplitude
as well as $W$--mediated box diagram of the SM.
One of them comes from another box diagrams 
containing $ W^{\prime}$
and others from tree level diagrams with
FCNC couplings involving $Z$ and $ Z^{\prime} $ bosons.
The relevant couplings to $W'$ bosons 
yielding leading contributions in the order of $\lambda$
are given by
\be
{\cal L}^{W^{\prime}} =  \frac{g \cot \phi}{\sqrt{2}} 
          \bar{t}_L \gamma^{\mu} 
(V_{td} d_L + V_{ts} s_L + V_{tb} b_L)
W^{\prime +}_{\mu} + \mbox{H.c.}~,
\ee
while the relevant FCNC couplings given by
\be
{\cal L}^Z_{_{FCNC}} & =& 
\frac{g}{2 \cos \theta} \lambda \sin^2 \phi
          \left[ V^{\ast}_{ts} V_{tb} \bar{s}_L \gamma^{\mu} b_L  +
                 V^{\ast}_{td} V_{tb} \bar{d}_L \gamma^{\mu} b_L
          \right] Z_{\mu} + \mbox{H.c.} 
\nonumber \\
{\cal L}^{Z^{\prime}}_{_{FCNC}}& =& \frac{g}{2 \sin \phi \cos \phi }
          \left[ V^{\ast}_{ts} V_{tb} \bar{s}_L \gamma^{\mu} b_L  +
                 V^{\ast}_{td} V_{tb} \bar{d}_L \gamma^{\mu} b_L
          \right] Z^{\prime}_{\mu} + \mbox{H.c.} ~. 
\ee
The new contribution by the box diagrams with one $W'$ boson
and with one $W$ boson is given by
\be
 M^{W'}_{12} = \frac{G^2_F m_{_{B_q}} \eta_{_{B_q}} m_{_{W}}^2}{3 \pi^2}
                 f^2_{B_q} B_{B_q} x_t f'_2(x_t,x_{_{W'}}) 
                 ( V_{tq} V^*_{tb} )^2 \cdot \cot^2 \phi~,
\ee
where $x_{_{W'}} = m_{_{W'}}^2/m_{_W}^2$ and
the function $f'_2(x)$ is given by
\be
{f'}_2(x,y) &=& \frac{1}{4y(x-y)^2(1-x)^2}
              \left[ (1-x)(4x^2+4y^2+5x^2y-8xy-4xy^2-x^2)
              \right.
\nonumber \\
          & &~~~~~~~~~~~~~~~~~
              \left.
          -3 x^2 (x-2y+xy) \log \left(\frac{y}{x} \right)
          -3 x (x-y)^2 \frac{\log y}{(y-1)}
              \right]~.
\ee
We can see that this function goes to $f_2(x)$ in Eq. (18)
when $y \to 1$.
Contributions from the $Z$-- and $Z'$--mediated FCNC interactions
are as follows:
\be
 M_{12}^{ Z} &=&  \frac{ \sqrt{2} G_F m_{_{B_q}} \eta_{_{B_q}} }
                       {12}
                  f^2_{B_q} B_{B_q} \cdot \lambda^2 \sin^4 \phi
                  ( V^{\ast}_{tq} V_{tb})^2 
\nonumber \\
 M_{12}^{ Z^{\prime}} &=& \frac{ \sqrt{2} G_F m_{_{B_q}} \eta_{_{B_q}}}
                               {12}
                  f^2_{B_q} B_{B_q} \cdot \lambda
                  ( V^{\ast}_{tq} V_{tb})^2~.
\ee
Note that the contribution from $Z'$--mediated FCNC is
of order of $\lambda$ while the other contribution 
from $Z$ boson are of order of $\lambda^2$.
Thus we expect that the $Z'$--mediated FCNC diagrams 
provide the dominant contribution among new physics effects.
We also note that all new physics contributions to $M_{12}^{(q)}$
involve the same CKM factors as those of the SM 
and has the same phase as the $W$--mediated box diagrams
in our choice of the bases for quarks.

We estimate each contributions 
comparing to the mass difference of the SM,
\be
\frac{\Delta M_{q}^{Z}}{\Delta M_q^W}
   &=& \lambda^2 \sin^4 \phi  \frac{\sqrt{2} \pi^2}{G_F m^2_{_W}}
                              \frac{1}{x_t f_2(x_t)}
   \simeq  71.9 \lambda^2 \sin^4 \phi \sim 5 \times 10^{-4}~, 
\nonumber \\
\frac{\Delta M_{q}^{Z^{\prime}}}{\Delta M_q^W}
   &=& \lambda \frac{\sqrt{2} \pi^2}{G_F m^2_{_W}} 
       \frac{1}{x_t f_2(x_t)}
       \simeq  71.9 \lambda  \sim 0.86 ~, 
\nonumber \\
\frac{\Delta M_q^{W^{\prime}}}{\Delta M_q^W} &=& 2 \cot^2 \phi
        \frac{f'_2(x_t,x_{_{W'}})}
             {f_2(x_t)} \simeq 2 \times 10^{-2}
\ee
where we have taken $|V_{tb}|$ =1 and 
$m_t$ = 175 Gev \cite{top}.  
To estimate numerical values, 
we use $\lambda = 0.012$ and $\sin^2 \phi = 0.22$
which give the central values of the measurement for 
the precision variables $\epsilon_1$ and $\epsilon_b$ 
obtained in Ref. \cite{lee}.
According to above results,
we find that mass differences normalized 
by the $W$--mediated box diagram 
are independent of the light quark type in the neutral $B$ mesons.
We also find that the $Z'$--mediated FCNC interactions dominates
and is comparable to the usual box diagram contribution.
The new physics contributions change 
the bounds of CKM matrix element 
$V_{td}$ and $V_{ts}$  as follows :
\begin{equation}
 0.08 < \left| \frac{V_{td}}{V_{cb}} \right| < 0.18 ~,
 ~ \left| \frac{V_{ts}}{V_{cb}} \right|  > 0.32 .
\end{equation}

As explained in the previous section, 
the CKM matrix of this model is no more unitary. 
In the case of $q=d$, 
the size of the unitarity violation is measured by
\be
V_{ud}^*V_{ub} + V_{cd}^*V_{cb} + V_{td}^*V_{tb} 
&=&  2 V_{td}^{0*} V^0_{tb} \cdot \lambda \sin^2 \phi 
\ee
which is of the linear order of $\lambda$.
With the values of $\lambda$ and $\sin^2 \phi$ used above, 
we obtain the bound for $|V_{td}|$:
\begin{equation}
 0.12 < \left| \frac{V_{td}}{V_{cb}} \right| < 0.33 ~,
\end{equation}
which is close to the unitary bound of the SM.
It is because the size of unitary violation is indeed small,
$ |V_{td}^{0*} V^0_{tb} \cdot \lambda \sin^2 \phi| \sim 10^{-6}$. 
Combining the unitarity bound with the bound from $B-\bar{B}$ mixing,
we could obtain the stronger bound on the value of $|V_{td}|$,
$ 0.12 < \left|V_{td}/V_{cb} \right| < 0.18$,
than that of the SM.

\section{ Rare Decays of $B$ Mesons }

Let us now examine the contributions of 
$Z$-- and $Z^{\prime}$--mediated FCNC processes 
to rare decays of neutral B mesons in topflavour model.
Considering the rare decays, we have to explore the uncertainties
of the predictions as well as 
the actual size of the branching ratios \cite{ali}.
New physics effects will be considered important in a particular
penguin decay only if they change the branching ratio
by quite a bit more than the uncertainty in the SM prediction. 

We are not interested in $b \to q \gamma$, $q=d,s$ decays
since this model does not alter the coupling with photon.
The annihilations  
$ B_{q}^0 \rightarrow l^+ l^- $, $q = d ,s$  
are quite interesting.
Although the decay rates of 
$ B_{q}^0 \rightarrow l^+ l^- $ have some hadronic uncertainties    
dependent upon the decay constant $f_{B_q}$,
they can be calculated rather precisely
since the renormalization--scale uncertainty is
quite small by including the QCD corrections \cite{buchala}.
The branching ratios are as follow \cite{ali}: 
\begin{eqnarray}
Br(B_{s}^0 \rightarrow \tau^+ \tau^- ) 
               &=& (7.4 \pm 2.1) \times 10^{-7} 
	~\left( \frac{f_{B_s}}{232~ \mbox{Mev}} \right)^2~, 
\nonumber \\
Br(B_{s}^0 \rightarrow \mu^+ \mu^- ) 
               &=& (3.5 \pm 1.0) \times 10^{-9} 
        ~\left( \frac{f_{B_s}}{ 232~ \mbox{Mev}} \right)^2 ~,
\nonumber \\
Br(B_{d}^0 \rightarrow \tau^+ \tau^- ) 
               &=& (3.1 \pm 2.9) \times 10^{-8} 
        ~\left(\frac{f_{B_d}}{ 200~ \mbox{Mev}} \right)^2~,
\\
Br(B_{d}^0 \rightarrow \mu^+ \mu^- ) 
               &=& (1.5 \pm 1.4) \times 10^{-10} 
        ~\left(\frac{f_{B_d}}{ 200~ \mbox{Mev}} \right)^2 ~,
\nonumber 
\end{eqnarray}
in the SM.
At present, the restrict experimental upper bounds are 
$Br(B_{s}^0 \rightarrow \mu^- \mu^+) < 8.4 \times 10^{-6} $  and 
$Br(B_{d}^0 \rightarrow \mu^- \mu^+) < 1.6 \times 10^{-6} $ \cite{cdf} 
and roughly four orders of magnitude larger than the SM predictions.  

In the topflavour model,
the decay processes due to
$Z$-- and $Z^{\prime}$--mediated FCNC couplings give
the amplitudes:
\begin{eqnarray}
{\cal M}_{Z} &=& - \frac{g^2}{4 m^2_{_Z} \cos^2 \theta} 
                 \lambda \sin^2 \phi   V^{0\ast}_{tq} V^0_{tb} 
		 f_{B_q} q^{\mu} \bar{l} \left[ g_L^l \gamma_{\mu} P_L 
		 + g_R^l \gamma_{\mu} P_R \right] l ~,
\nonumber \\
{\cal M}_{Z^{\prime}} &=& - \frac{g^2}{4 m^2_{_{Z^{\prime}}}}  
                 \frac{1}{\sin \phi \cos \phi}
		 V^{0\ast}_{tq} V^0_{tb} 
		 f_{B_q} q^{\mu} \bar{l}
		 \left[ g_L^{\prime l} \gamma_{\mu} P_L 
		 + g_R^{\prime l} \gamma_{\mu} P_R \right] l ~,
\end{eqnarray}
where the shifted couplings are given by
\begin{eqnarray}
g_L^e &=& g_L^{\mu} = (-\frac{1}{2} + \sin^2 \theta ) 
                      + \frac{\lambda}{2} \sin^4 \phi ~,
\nonumber \\
g_L^{\tau} &=& (-\frac{1}{2} + \sin^2 \theta) 
               - \frac{\lambda}{2} \sin^2 \phi \cos^2 \phi~,
\nonumber \\
g_R^l &=& \sin^2 \theta ~,
~~~~~~ l = e ,\mu , \tau ,
\end{eqnarray}
and the couplings to $Z'$ are
\begin{eqnarray}
g_L^{\prime e}&=& g_R^{\prime l} = -\frac{\sin \phi}{2 \cos \phi} - 
                        \frac{\lambda}{2} 
			\frac{\sin^3 \phi \cos \phi}{\cos^2 \theta}
			( 1 - 2 \sin^2 \theta)~, 
\nonumber \\
g_L^{\prime \tau} &=& \frac{\cos \phi}{2 \sin \phi} - 
                        \frac{\lambda}{2} 
			\frac{\sin^3 \phi \cos \phi}{\cos^2 \theta}
			( 1 - 2 \sin^2 \theta)~, 
\nonumber \\
g_R^{\prime l} &=& -\lambda \sin^3 \phi \cos \phi \tan^2 \theta ~,
~~~~~~ l = e ,\mu , \tau~. 
\end{eqnarray}
We obtain the branching ratios of rare decays due to
FCNC interactions of Eq. (29) as
\begin{eqnarray}
Br( B_s^0 \rightarrow  \tau^+ \tau^-)_{FCNC}  &\simeq& 
   9.5  \times10^{-7}  
   \left( \frac{f^2_{B_s}}{232~ \mbox{Mev}}\right)^2 ~, 
\nonumber \\
Br( B_s^0 \rightarrow \mu^+ \mu^-)_{FCNC} &\simeq&  
   1.2 \times 10^{-9}  
   \left( \frac{f^2_{B_s}}{232~ \mbox{Mev}}\right)^2 ~, 
\nonumber \\
Br( B_d^0 \rightarrow  \tau^+ \tau^-)_{FCNC}  &\simeq& 
   6.5  \times10^{-8}  
   \left( \frac{f^2_{B_d}}{200~ \mbox{Mev}} \right)^2~, 
\\ 
Br( B_d^0 \rightarrow  \mu^+ \mu^-)_{FCNC}  &\simeq& 
   8.2  \times10^{-11}  
   \left( \frac{f^2_{B_d}}{200~ \mbox{Mev}}\right)^2 ~, 
\nonumber 
\end{eqnarray}
where the values of $\lambda$ and $\sin^2 \phi$ are
taken as used in the previous section.
We find that the contributions to the decays into $\mu$ pair 
is less than those of the SM predictions,
while the contributions to the decays into $\tau$ pair
are comparable with the SM predictions.
This is caused by the fact that the couplings of
$\tau$ pair to $Z$ and $Z'$ bosons are stronger
than those of $e$ and $\mu$ pairs.
The enhancement of the decay rate of 
$B_s^0 \to \tau^+ \tau^-$ can be observed 
since the total branching ratio will be 
four times of the theoretical uncertainty 
higher from of the SM prediction.
However the decay rates into $B_d^0 \to \tau^+ \tau^-$ are
at most two times of the theoretical uncertainty higher and 
consequently this can only be considered 
marginal signals of new physics at best.

The gluon--mediated hadronic decays 
and $b \to q l^+ l^-$ processes
have larger errors than the annihilation decays \cite{ali}
and it is hard to expect the enhancement 
of the effects of the topflavour model by order of magnitudes 
since new contributions of this model always 
have the same CKM factors as those of the SM.
Thus we hardly expect to find the new physics signals
in the gluon--mediated hadronic decays and $b \to q l^+ l^-$ decays,
and we do not show the explicit calculations 
for those processes here.

\section{Discussions and Conclusion}

We study the effects of the topflavour model 
on the $B-\bar{B}$ mixings and rare decays.
In the generic model containing the FCNC interactions
at tree level \cite{fcnc},
the rare decays are much enhanced by one or more order
of magnitudes due to the FCNC interactions
when the new physics effect on the $B-\bar{B}$ mixings are 
observable \cite{gronau}. 
However we find the sizable contributions of this model to
the $B_d-\bar{B}_d$ mixings while there is no "smoking gun"
signals in rare $B_d^0$ decays.
In fact, the FCNC coupling of this model itself is too
small to produce significant contributions to decay rates.
As estimated in Ref. \cite{gronau}, 
the FCNC couplings have to be of order of 10$^{-3}$
to give substantial enhancements of branching ratios
of rare decays enough for observation.
In the topflavour model, the couplings are suppressed
by the CKM matrix elements $|V_{td}| \sim 10^{-3}$ or
$|V_{ts}| \sim 10^{-2}$ as well as
the model parameters $\lambda \sim 10^{-2}$ 
and $\sin^2 \phi \sim 10^{-1}$.
The coupling to $Z'$ boson is not suppressed by $\sin^2 \phi$
but still of order of 10$^{-5}$.
In the $B-\bar{B}$ mixing amplitude, the contribution of
$Z'$ exchange diagram is rather large and 
even comparable to the SM contribution.
Contribution of this diagram is of order of $\lambda$ while 
that of the $Z$ exchange diagram is of order of $\lambda^2$.
And the FCNC coupling to $Z'$ is not suppressed by
$\sin^2 \phi$.
(Note that both contributions to the branching ratios
of rare decays are always of order of $\lambda^2$.)
Meanwhile the $W'$ mediated box diagrams is of order of $\lambda$,
but suppressed by loop integrals.

We showed that the bound of $|V_{td}|$ from the $B_d-\bar{B}_d$ mixing
can be considerably changed in the topflavour model
with the value of $\lambda$ and $\sin \phi$ fitted by the LEP data.
Therefore when we directly measure the very low value of $|V_{td}|$ 
in the future and rare decay rates of $B_d^0$ 
are still consistent with the SM predictions,
it would be a critical test of the topflavour model.

Most of the experimental bounds for flavour--changing processes
are not so strong that cannot further constrain this model.
However, we can find interesting possibility.
If we have the better experimental bound on $|V_{td}|$
from the $B_d-\bar{B}_d$ mixing, it is possible that 
the value of $|V_{td}|$ could not satisfy 
the $B_d-\bar{B}_d$ mixing bound 
and the unitarity bound at the same time 
in some region of ($\lambda$, $\sin^2 \phi$) parameter space
because the $B_d-\bar{B}_d$ mixing bound
is shifted from that of the SM in this model.
Then we will have an additional constraint on the model parameters
to avoid such contradiction.

Finally we wish to remark the general treatment
on the quark mixing matrix.
In this paper, all couplings to the charged currents 
not including $t$ quark contain the common factor 
$(1 - \lambda \sin^4 \phi)$ of which effect is washed out 
by introducing the Fermi constant defined by $\mu$ decay.
When we take general unitary matrices for $V_U$ and $V_D$,
each charged current coupling is differently changed,
which depends on the elements ${V_U}_{ij}$ and ${V_D}_{ij}$.
If so, we have more undetermined parameters and
we should restrict them by many low--energy processes 
such as $\beta$--decay, $\pi$ decays, $K$ decays etc..
But it is still hard to define the CKM matrix in that case.
More interestingly additional phase is introduced 
from $V_U$ in the general case. 
It provides many new possibilities of CP violating phenomena
which is not tested yet.

In conclusion we considered the effects of the topflavour model
in the $B_d-\bar{B}_d$ mixings and rare $B$ decays.
We found that the $B_d-\bar{B}_d$ mixings are substantially 
affected but rare decays of $B_d$ are not 
because the effects on the $B_d-\bar{B}_d$ mixings are
of order of $\lambda$ while those on the rare decays
of $\lambda^2$.
This may be a characteristic feature of this model 
while other new physics containing tree level FCNC interactions
do not have.

\vskip 0.8cm
\begin{center}
{\bf Acknowledgement}
\end{center}
\vskip 0.6cm

We thank Prof. Ko for careful reading of the manuscript
and valuable comments.
This work was supported in part by 
KAIST Center for Theoretical Physics and Chemistry  (K. Y. L.) and
in part by the Korean Science and Engineering Foundation (KOSEF).


\vskip 2.6cm


\begin{references}

\bibitem{ua1} C. Albajar {\it et al.}, UA1 collaboration,
Phys. Lett. B {\bf 186}, 247 (1987).

\bibitem{argus} H. Albrecht {\it et al.}, ARGUS collaboration,
Phys. Lett. B {\bf 192}, 245 (1987).

\bibitem{deltam} L. Gibbons (CLEO collaboration),
Invited talk at the International Conference on High Energy Physics,
Warsaw, Poland, ICHEP 96 (1996); 
ALEPH collaboration, Contributed paper to ICHEP 96 (1996).

\bibitem{alilondon} A. Ali and D. London,
DESY Report No. DESY 96--140, hep-ph/9607392.

\bibitem{malkawi} E. Malkawi, T. Tait and C.-P. Yuan,
Phys. Lett. B {\bf 385}, 304 (1996).

\bibitem{nandi} D. J. Muller and S. Nandi,
Phys. Lett. B {\bf 383}, 345 (1996).

\bibitem{lee} J. C. Lee, K. Y. Lee and J. K. Kim,
Report No. KAIST-TH-97/19, hep-ph/9711509, 
to appear in Phys. Lett. B.

\bibitem{ETC} R. S. Chivukula, E. H. Simmons and J. Terning,
Phys. Lett. B {\bf 331}, 383 (1994);
Phys. Rev. D {\bf 53}, 5258 (1996).

\bibitem{smbb} J. Hagelin, Nucl. Phys. B {\bf 193}, 123 (1981);
T. Inami and C. S. Lim, Prog. Theo. Phys. {\bf 65}, 297 (1981);
{\bf 65}, 772 (E) (1982);
A. Buras, W. Slominski and H. Steger, 
Nucl. Phys. B {\bf 238}, 529 (1984); B {\bf 245}, 369 (1984).

\bibitem{top} D. S. Kestenbaum, Report No. FERMILAB-CONF-97-016,
presented at the 16th International Conference 
on Physics in Collision, Mexico (1996).

\bibitem{ali} {\sl For a review, see} A. Ali,
DESY Report No. DESY 96--106, hep-ph/9606324 and references therein.

\bibitem{buchala} G. Buchala and A. J. Buras, 
Nucl. Phys. B {\bf 400}, 225 (1993).

\bibitem{cdf} F. Abe {\it et al.}, CDF collaboration,
Phys. Rev. Lett. {\bf 76}, 4675 (1996).

\bibitem{fcnc} Y. Nir and D. Silverman, 
Phys. Rev. D {\bf 42}, 1477 (1990);
D. Silverman, {\it ibid} {\bf 45}, 1800 (1992);
G. C. Branco, T. Morozumi, P. A. Parada amd M. N. Rebelo, 
{\it ibid} {\bf 48}, 1167 (1993);
V. Barger, M. S. Berger and R. J. N. Phillips,
{\it ibid} {\bf 52}, 1663 (1995).

\bibitem{gronau} M. Gronau and D. London,
Phys. Rev. D {\bf 55}, 2845 (1997).


\end{references}
\end{document}